\documentclass[twocolumn,prd,showpacs,nofootinbib,superscriptaddress,preprintnumbers,floatfix]{revtex4}

\usepackage{amsmath}
\usepackage{amsfonts}
\usepackage{amssymb}
\usepackage[x11names]{xcolor}
\usepackage{epsfig}
\usepackage{graphicx}
\usepackage{nicefrac}
\usepackage{hhline}

\definecolor{markus}{rgb}{0.69,0.65,0.43}
\definecolor{heidelbeer}{rgb}{0.5,0,0.5}
\newcommand{\I}{\text{i}}
\newcommand{\E}{\text{e}}

\newcommand{\Eqref}[1]{Eq.~\eqref{#1}}

\newcommand{\te}{\tilde{e}}
\newcommand{\ta}{\tilde{\alpha}}
\newcommand{\tm}{\tilde{m}}

\newcommand{\Fd}{\widetilde{F}}

\begin{document}

\preprint{DESY 06-221; DCPT/06/182; IPPP/06/91}

\title {
  On the Particle Interpretation of the PVLAS Data:\\
  Neutral versus Charged Particles 
}

\author{Markus Ahlers}
\email{markus.ahlers@desy.de}
\affiliation{Deutsches Elektronen-Synchrotron DESY, Notkestra\ss e 85, D-22607 Hamburg, Germany}
\author{Holger Gies}
\email{h.gies@thphys.uni-heidelberg.de}
\affiliation{Institut f\"ur Theoretische Physik, Universit\"at Heidelberg, Philosophenweg 16,
D-69120 Heidelberg, Germany}
\author{Joerg Jaeckel}
\email{joerg.jaeckel@durham.ac.uk}
\affiliation{Centre for Particle Theory, Durham University, Durham, DH1 3LE, UK}
\author{Andreas Ringwald}
\email{andreas.ringwald@desy.de}
\affiliation{Deutsches Elektronen-Synchrotron DESY, Notkestra\ss e 85, D-22607 Hamburg, Germany}

\begin{abstract}
Recently the PVLAS collaboration reported the observation of a
rotation of linearly polarized laser light induced by a transverse
magnetic field - a signal being unexpected within standard QED.  Two
mechanisms have been proposed to explain this result: production of a
single (pseudo-)scalar particle coupled to two photons or pair
production of light millicharged particles.  In this work, we study
how the different scenarios can be distinguished.  We summarize the
expected signals for vacuum magnetic dichroism (rotation) and
birefringence (ellipticity) for the different types of particles -
including new results for the case of millicharged scalars.  The sign
of the rotation and ellipticity signals as well as their dependencies
on experimental parameters, such as the strength of the magnetic field
and the wavelength of the laser, can be used to obtain information
about the quantum numbers of the particle candidates and to
discriminate between the different scenarios.  We perform a
statistical analysis of all available data resulting in strongly
restricted regions in the parameter space of all scenarios. These
regions suggest clear target regions for upcoming experimental
tests. As an illustration, we use preliminary PVLAS data to
demonstrate that near future data may already rule out some of these
scenarios.
\end{abstract}


\maketitle

\section{\label{intro} Introduction}

The absorption probability and the propagation speed of polarized
light propagating in a magnetic field depends on the
relative orientation between the polarization and the magnetic field.
These effects are known as vacuum magnetic dichroism and birefringence,
respectively, resulting from fluctuation-induced vacuum polarization.

In a pioneering experiment, the BFRT collaboration searched for these
effects by shining linearly polarized laser photons through a
superconducting dipole magnet.  No significant signal was found, and a
corresponding upper limit was placed on the rotation (dichroism) and
ellipticity (birefringence) of the photon beam developed after passage
through the magnetic field~\cite{Semertzidis:1990qc,Cameron:1993mr}.

Recently, however, a follow-up experiment done by the PVLAS
collaboration reported the observation of a rotation of the
polarization plane of light after its passage through a transverse
magnetic field in vacuum~\cite{Zavattini:2005tm}. Moreover,
preliminary results presented by the PVLAS collaboration at various
seminars and conferences hint also at the observation of an
ellipticity (birefringence)~\cite{PVLASICHEP,Cantatore:IDM2006}.

These findings have initiated a number of theoretical and experimental
activities, since the magnitude of the reported signals exceeds the
standard-model expectations by far.\footnote{The
incompatibility with standard QED has recently been confirmed again in
a more careful wave-propagation study which also takes the rotation of
the magnetic field in the PVLAS setup properly into account
\cite{Adler:2006zs,Biswas:2006cr}. The proposal of a potential QED
effect in the rotating magnetic field \cite{Mendonca:2006pg} is
therefore ruled out.} If the observed effects are indeed true signals
of vacuum magnetic dichroism and birefringence and not due to a
subtle, yet unidentified systematic effect, they signal new physics
beyond the standard model of particle physics.

One obvious possible explanation, and indeed the one which was also a
motivation for the BFRT and PVLAS experiments, may be offered by the
existence of a new light neutral spin-$0$ boson
$\phi$~\cite{Maiani:1986md}.  In fact, this possibility has been
studied in Ref.~\cite{Zavattini:2005tm}, with the conclusion that the
rotation observed by PVLAS can be reconciled with the non-observation
of a rotation and ellipticity by BFRT, if the hypothetical neutral
boson has a mass in the range $m_\phi\sim (1-1.5)$~meV and a coupling
to two photons in the range 
$g\sim (1.7-5.0)\times 10^{-6}$~GeV$^{-1}$. 

Clearly, these values almost certainly exclude the possibility that
$\phi$ is a genuine QCD axion $A$~\cite{Weinberg:1977ma,%
Wilczek:1977pj}.  For the latter, a mass $m_A\sim 1$\,meV implies a
Peccei-Quinn symmetry~\cite{Peccei:1977hh,Peccei:1977ur} breaking
scale $f_A\sim 6\times 10^{9}$\,GeV. 
Since, for an axion, $g\sim \alpha
|E/N|/(2\pi f_A)$~\cite{Bardeen:1977bd,Kaplan:1985dv,Srednicki:1985xd}, 
one would need an extremely large ratio $|E/N|\sim
3\times 10^7$ of electromagnetic and color anomalies in order to
arrive at an axion-photon coupling in the range suggested by PVLAS.
This is far away from the predictions of any model conceived so far.
Moreover, such a new, axion-like particle (ALP) must have very
peculiar properties~\cite{Masso:2005ym,Jain:2005nh,%
Jaeckel:2006id,Masso:2006gc,Mohapatra:2006pv,Jain:2006ki} in order to
evade the strong constraints on its two photon coupling from stellar
energy loss considerations~\cite{Raffelt:1996} and from its
non-observation in helioscopes such as the CERN Axion Solar Telescope
(CAST)~\cite{Zioutas:2004hi}. A light scalar boson is
furthermore constrained by upper limits on non-Newtonian forces
\cite{Dupays:2006dp}.

Recently, an alternative to the ALP interpretation of the PVLAS
results was proposed~\cite{Gies:2006ca}.  It is based on the
observation that the photon-initiated real and virtual pair production
of millicharged particles (MCPs) $\epsilon^\pm$ in an external
magnetic field would also manifest itself as a vacuum magnetic
dichroism and ellipticity.  In particular, it was pointed out that the
dichroism observed by PVLAS may be compatible with the non-observation
of a dichroism and ellipticity by BFRT, if the millicharged particles
have a small mass $m_\epsilon\sim 0.1$~{\rm eV} and a tiny fractional
electric charge 
$\epsilon\equiv Q_\epsilon/e \sim 10^{-6}$.  
As has been shown recently~\cite{Masso:2006gc}, such particles may be
consistent with astrophysical and cosmological bounds (for a review,
see Ref.~\cite{Davidson:2000hf}), if their tiny charge arises from
gauge kinetic mixing of the standard model hypercharge U(1) with
additional U(1) gauge factors from physics beyond the standard
model~\cite{Holdom:1985ag}.  This appears to occur quite naturally
in string theory~\cite{Abel:2006qt}.

It is very comforting that a number of laboratory-based {low-energy}
tests of the ALP and MCP interpretation of the PVLAS anomaly are
currently set up and expected to yield decisive results within the
upcoming year. For instance, the Q\&A experiment has very
recently released first rotation data \cite{Chen:2006cd}. Whereas the
Q\&A experimental setup is qualitatively similar to PVLAS, the
experiment operates in a slightly different parameter region; here, no
anomalous signal has been detected so far. 

\begin{figure}
{\centering\includegraphics[width=8.5cm]{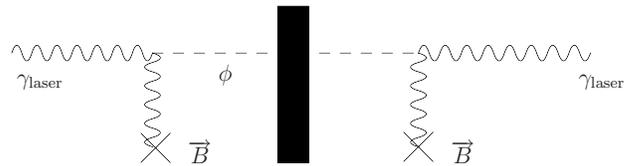}}
\caption{Schematic view of a ``light shining through a wall'' experiment.
(Pseudo-)scalar production through photon conversion in a magnetic field (left), subsequent travel
through a wall, and final detection through photon regeneration
(right). }
\label{fig:ph_reg}
\end{figure}

The interpretation of the PVLAS signal involving an ALP that
interacts weakly with matter will crucially be tested by photon
regeneration (sometimes called ``light shining through walls'')
experiments~\cite{Sikivie:1983ip,Anselm:1986gz,%
Gasperini:1987da,VanBibber:1987rq,Ruoso:1992nx,Ringwald:2003ns,%
Gastaldi:2006fh} presently 
under construction or serious consideration~\cite{Pugnat:2005nk,Rabadan:2005dm,%
Cantatore:Patras,Kotz:2006bw,Baker:Patras,Rizzo:Patras,ALPS}. 
In these experiments (cf. Fig.~\ref{fig:ph_reg}), a photon beam is shone across
a magnetic field, where a fraction of them turns into ALPs. The ALP
beam can then propagate freely through a wall or another obstruction
without being absorbed, and finally another magnetic field located on
the other side of the wall can transform some of these ALPs into
photons --- {seemingly} regenerating these photons out of nothing.
Another probe could be provided by direct astrophysical
observations of light rays traversing a pulsar magnetosphere in binary
pulsar systems~\cite{Dupays:2005xs}.

Clearly, photon regeneration will be negligible for MCPs. 
Their existence, however, 
can be tested by improving
the sensitivity of instruments for the detection of vacuum magnetic
birefringence and 
dichroism~\cite{Cameron:1993mr,Zavattini:2005tm,%
  Chen:2006cd,Rizzo:Patras,Pugnat:2005nk,Heinzl:2006xc}.
Another sensitive tool is Schwinger pair production in strong electric
fields, as they 
are available,
for example, in accelerator
cavities~\cite{Gies:2006hv}. A classical probe for MCPs is the search
for invisible orthopositronium
decays~\cite{Dobroliubov:1989mr,Mitsui:1993ha}, for which new
experiments are currently running~\cite{Badertscher:2006fm} or being
developed~\cite{Rubbia:2004ix,Vetter:2004fs}.

From a theoretical perspective, the two scenarios are substantially
different: the ALP scenario is parameterized by an effective
non-renormalizable dimension-5 operator, the stabilization of which
almost inevitably requires an underlying theory at a comparatively low
scale, say in between the electroweak and the GUT scale. By contrast,
the MCP scenario in its simplest version is reminiscent to QED; it is
perturbatively renormalizable and can remain a stable microscopic
theory over a wide range of scales.

{The present paper is devoted to an investigation of the
  characteristic properties of the different scenarios in the light of
  all available data collected so far. A careful study of the optical
  properties of the magnetized vacuum can indeed reveal important
  information about masses, couplings and other quantum numbers of the
  potentially involved hypothetical particles. This is quantitatively
  demonstrated by global fits to all published data. For further
  illustrative purposes, we also present global fits which include the
  preliminary data made available by the PVLAS collaboration at
  workshops and conferences. We stress that this data is only used
  here to qualitatively demonstrate how the optical measurements can
  be associated with particle-physics properties. Definite
  quantitative predictions have to await the outcome of a currently
  performed detailed data analysis of the PVLAS collaboration. Still,
  the} resulting fit regions can be viewed as a {preliminary
  estimate of} ``target regions'' for the various laboratory tests
mentioned above. Moreover, the statistical analysis is also meant to
help the theorists in deciding whether they should care at all about
the PVLAS anomaly, and, if yes, whether {there is a
  pre-selection of phenomenological models or model building blocks
  that deserve to be studied in more detail. }

The paper is organized as follows. In the next section \ref{sec2} we
summarize the signals for vacuum magnetic dichroism and birefringence
in presence of axion-like and millicharged particles.  We use these
results in Sec.~\ref{sec3} to show how the different scenarios
can be distinguished from each other and how information about the
quantum numbers of the potential particle candidates can be
collected.  In Sec.~\ref{sec4} we then perform a statistical analysis
including all current data. We also use preliminary PVLAS data to show
the prospects for the near future.  We summarize our conclusions  in
Sec.~\ref{conclusions}.

\section{ Vacuum Magnetic Dichroism, Birefringence, and Photon
    Regeneration}\label{sec2}

We start here with some general kinematic considerations relevant to
dichroism and birefringence, which are equally valid for the case of
ALP  and the case of MCP production.

Let $\vec k$ be the momentum of the incoming photon, with $|\vec
k|=\omega$, and let $\vec B$ be a static homogeneous magnetic field,
which is perpendicular to $\vec k$, as it is the case in all of the
{afore}-mentioned polarization experiments.

The photon-initiated production of an  ALP  with mass $m_\phi$ or an MCP
with mass $m_\epsilon$, leads, for $\omega > m_\phi$ or $\omega > 2
m_\epsilon$, respectively, to a  non-trivial ratio of the
survival probabilities $\exp(-\pi_{\parallel,\perp}(\ell))$ of a
photon  after it has traveled a distance $\ell$, for
photons polarized parallel $\parallel$ or perpendicular $\perp$ to
$\vec B$.
This  non-trivial ratio  manifests itself directly in a
dichroism: for a linearly polarized photon beam, the angle $\theta$
between the initial polarization vector and the magnetic field will
change to $\theta + \Delta \theta$ after passing a distance $\ell$
through the magnetic field, with
\begin{equation}
\cot (\theta+\Delta\theta)=\frac{E_{\parallel}}{E_{\perp}}
=\frac{E^{0}_{\parallel}}{E^{0}_{\perp}}
\exp\left(-\frac{1}{2}(\pi_{\parallel}(\ell)
                       -\pi_{\perp}(\ell))\right).
\label{eq1}
\end{equation}
Here, $E_{\parallel,\perp}$ are the electric field components of the
laser parallel and perpendicular to the external
magnetic field, and the superscript ``0'' denotes initial values.
For small rotation angle $\Delta\theta$, we have

\begin{equation}
\label{delthet}
\Delta\theta \simeq \frac{1}{4}(\pi_{\parallel}
    -\pi_{\perp})\, \sin(2\theta).
\end{equation}
We will present the results for the probability exponents
$\pi_{\parallel}-\pi_{\perp}$ for  ALPs  and MCPs in the
following subsections.

Let us now turn to birefringence. The propagation speed of the laser
photons is slightly changed in the magnetic field owing to the
coupling to virtual  ALPs   or 
MCPs.
Accordingly, 
the time $\tau_{\parallel,\perp}(\ell)$ it takes
for a photon to traverse a distance $\ell$ differs for the two
polarization modes, causing a phase difference between the two modes,
\begin{equation}
\Delta\phi=\omega (\tau_{\parallel}(\ell)
-\tau_{\perp}(\ell)).
\end{equation}
This induces an ellipticity $\psi$ of the outgoing beam,
\begin{equation}
\label{psi}
\psi =\frac{\omega}{2}(\tau_{\parallel}(\ell)
    -\tau_{\perp}(\ell))\sin(2\theta),
\quad\quad\rm{for}\,\,\psi\ll1.
\end{equation}
Again, we will present the results for
$\tau_{\parallel}-\tau_{\perp}$
for  ALPs  and MCPs in the following subsections. \bigskip

\subsection{Production of Neutral Spin-0 Bosons}

A neutral spin-0 particle can interact with two photons via
\begin{equation}
{\mathcal  L}^{(+)}_{\rm{int}}
  =-\frac{1}{4}g\phi^{(+)}F_{\mu\nu}F^{\mu\nu}
  =\frac{1}{2}g\phi^{(+)}(\vec{E}^{2}-\vec{B}^{2}),
\end{equation}
if it is a scalar, or
\begin{equation}
{\mathcal L}^{(-)}_{\rm{int}}
  =-\frac{1}{4}g\phi^{(-)}F_{\mu\nu}\widetilde{F}^{\mu\nu}
  =g\phi^{(-)}(\vec{E}\cdot\vec{B}),
\end{equation}
if it is a pseudoscalar. In a homogeneous magnetic background  $\vec{B}$, the
leading order contribution to the conversion (left half of Fig.
\ref{fig:ph_reg}) of (pseudo-)scalars into photons comes from the
terms $\sim \vec{B}^2$ and $\sim \vec{E}\cdot\vec{B}$, respectively.
The polarization of a photon is now given by the direction of the
electric field of the photon, $\vec{E}_\gamma$, whereas its
magnetic field, $\vec{B}_{\gamma}$ is perpedicular to the
polarization. Therefore, only those fields polarized perpendicular
(parallel) to the background magnetic field will have nonvanishing
$\vec{B}_{\gamma}\cdot\vec{B}\neq0$
($\vec{E}_{\gamma}\cdot\vec{B}\neq0$) and interact with the
\mbox{(pseudo-)scalar} particles. Accordingly, for scalars we have,
\begin{equation}
\pi^{(+)}_{\perp}\neq 0,\quad \pi^{(+)}_{\parallel}=0,\quad
\tau^{(+)}_{\perp}\neq 0, \quad \tau^{(+)}_{\parallel}=0
\end{equation}
whereas for pseudoscalars we find
\begin{equation}
\pi^{(-)}_{\perp}= 0,\quad \pi^{(-)}_{\parallel}\neq
  0,\quad
\tau^{(-)}_{\perp}= 0, \quad \tau^{(-)}_{\parallel}\neq 0.
\end{equation}
Apart from this, the interaction is identical  in lowest order,
\begin{equation}
 \pi^{(+)}_{\perp}=\pi^{(-)}_{\parallel}\,\,{\rm{and}}\,\,
\tau^{(+)}_{\perp}=\tau^{(-)}_{\parallel}.
\end{equation}
Using Eqs.~\eqref{eq1}-\eqref{psi} we deduce
\begin{equation}
{{\Delta\theta}}^{(+)}
  =-{{\Delta\theta}}^{(-)},\,\,
{\rm{and}}\,\,\psi^{(+)}=-\psi^{(-)}.
\end{equation}

We can now summarize the predictions on the rotation
${{\Delta\theta}}$ and the ellipticity $\psi$ in (pseudo-)scalar ALP
models with coupling $g$ and mass  $m_\phi$
\cite{Maiani:1986md,Raffelt:1987im}.  We assume a setup
as in the BFRT experiment with a dipole magnet of length $L$ and
homogeneous magnetic field $B$. The polarization of the laser beam
with photon energy $\omega$ has an angle $\theta$ relative to the
magnetic field. The effective number of passes of photons in the
dipole is $N_\text{pass}$.  Due to coherence, the
rotation ${{\Delta\theta}}$ and ellipticity $\psi$
depend non-linearly on the length of the apparatus $L$ and linearly on
the number of passes $N_{\rm{pass}}$, instead of simply being
proportional to $\ell=N_{\rm pass}L$; whereas the photon component
is reflected at the cavity mirrors, the ALP component is not and
leaves the cavity after each pass:
\begin{equation}
\label{dicALP}
-{{\Delta\theta}}^{(+)} ={{\Delta\theta}}^{(-)}
  = N_\text{pass}\left(\frac{gB\omega}{m_\phi^2}\right)^2
    \sin^2\left(\frac{L m_\phi^2}{4\omega}\right)\sin2\theta,
\end{equation}
\begin{eqnarray}
\label{birALP}
-\psi^{(+)} \!\!&=&\!\!\psi^{(-)}
\\\nonumber
\!\!&=&\!\! \frac{N_\text{pass}}{2}\left(\frac{gB\omega}{m_\phi^2}\right)^2\left(\frac{L m_\phi^2}{2\omega}-
\sin\left(\frac{L m_\phi^2}{2\omega}\right)\right)\sin2\theta.
\end{eqnarray}

For completeness, we present here also the flux of regenerated photons
in a ``light-shining through a wall''
experiment (cf. Fig.~\ref{fig:ph_reg}). In the case of a pseudoscalar,
it reads
\begin{equation}
\label{regALPps}
\dot{N}_{\gamma\ {\rm reg}}^{(-)} = \dot{N_0}\left\lfloor
\frac{N_\text{pass}+1}{2}\right\rfloor
\frac{1}{16}\left(gBL\cos\theta\right)^4\left(
\frac{\sin(\frac{L m_\phi^2}{4\omega})}
{\frac{L m_\phi^2}{4\omega}}
\right)^4,
\end{equation}
where $\dot{N}_0$ is the original photon flux. For a scalar,
the $\cos\theta$ is replaced by a $\sin\theta$.
Equation (\ref{regALPps}) is for the special situation in which a dipole of length $L$ and field $\vec{B}$ is
used for generation as well as for regeneration of the ALPs as it is the case for the BFRT experiment.
Note that only passes towards the wall count.

\subsection{Optical Vacuum Properties from Charged-Particle
  Fluctuations}

Let us now consider the interactions between the laser beam and the
magnetic field mediated by fluctuations of particles with charge
$\epsilon e$ and mass $m_\epsilon$. For laser frequencies above
threshold, $\omega>2m_\epsilon$, pair production becomes possible in
the magnetic field, resulting in a depletion of the incoming photon
amplitude. The corresponding photon attenuation coefficients
$\kappa_{\|,\bot}$ for the two polarization modes are related to the
probability exponents $\pi_{\|,\bot}$ by
\begin{equation}
\pi_{\|,\bot}=\kappa_{\|,\bot}\, \ell, \label{pikappa}
\end{equation}
depending linearly on the optical path length $\ell$. Also the time
$\tau_{\|,\bot}$ it takes for the photon to traverse the interaction
region with the magnetic field exhibits the same dependence,
\begin{equation}
\tau_{\|,\bot}=n_{\|,\bot}\, \ell, \label{taun}
\end{equation}
where $n_{\|,\bot}$ denotes the refractive indices of the magnetized
vacuum.

\subsubsection{Dirac Fermions}
\label{sec:MCF}

We begin with vacuum polarization and pair production of
  charged Dirac fermions~\cite{Gies:2006ca}, arising from an
  interaction Lagrangian
\begin{equation}
\label{lintDsp}
{\mathcal L}_{\rm int}^{\rm Dsp}= \epsilon\, e\,
\overline{\psi}_{\epsilon}
\gamma_\mu \psi_{\epsilon} A^\mu
,
\end{equation}
with $\psi_\epsilon$ being a Dirac spinor {(``Dsp'')}.

Explicit expressions for the photon absorption coefficients
$\kappa_{\parallel,\perp}$ can be inferred from the polarization
tensor which is obtained by integrating over the fluctuations of the
$\psi_\epsilon$ field. This process $\gamma\to \epsilon^+\epsilon^-$
has been studied frequently in the literature for the case of a
homogeneous magnetic
field~\cite{Toll:1952,Klepikov:1954,Erber:1966vv,Baier:1967,Klein:1968,Adler:1971wn,Tsai:1974fa,%
Daugherty:1984tr,Dittrich:2000zu}: 
\begin{align}\label{absorption}
\pi_{\|,\bot}^{\rm Dsp} &\equiv\kappa_{\parallel,\perp}^{\rm Dsp}\ell
= \frac{1}{2}\epsilon^3 e \alpha \frac{B \ell }{m_\epsilon}\,
T_{\parallel,\perp}^{\rm Dsp}(\chi )
\\[1.5ex]
\nonumber
&= 1.09\times 10^6\ \epsilon^3
\left( \frac{\rm eV}{m_\epsilon} \right)
\left( \frac{B}{\rm T}\right)
\left( \frac{\ell}{\rm m}\right)\,T_{\parallel,\perp}^{\rm Dsp}(\chi )
,
\end{align}
where $\alpha = e^2/4\pi$ is the fine-structure constant.  Here,
$T_{\parallel,\perp }^{\rm Dsp}(\chi )$ has the form of a parametric
integral~\cite{Tsai:1974fa},
%
\begin{multline}
\label{absorb}
T_{\parallel,\perp}^{\rm Dsp} =
\frac{4\sqrt{3}}{\pi\chi}
\int\limits_0^1 {\rm d}v\
K_{2/3}\left( \frac{4}{\chi}\frac{1}{1-v^2}\right)
\\
\times
\frac{\left[ \left( 1-\frac{1}{3}v^2\right)_\parallel,
\left(\frac{1}{2} +\frac{1}{6}v^2\right)_\perp
\right]}{(1-v^2)}
\end{multline}
\begin{flushright}
$\displaystyle
= \begin{cases}
\sqrt{\frac{3}{2}}\ {\rm e}^{-4/\chi}\ \left[(\frac{1}{2})_\parallel,(\frac{1}{4})_\perp\right] & \text{for}
\,\,\chi\ll 1\,\,\text{,} \\
\frac{2\pi}{\Gamma\left(\frac{1}{6}\right)\Gamma\left(\frac{13}{6}\right)}
\chi^{-1/3}\left[ (1)_\parallel,(\frac{2}{3})_\perp\right]& \text{for}  \,\,\chi\gg 1\,\,\text{,}
\end{cases}
$
\end{flushright}
the dimensionless parameter $\chi$ being defined as
\begin{equation}
\label{chi}
\chi \equiv  \frac{3}{2} \frac{\omega}{m_\epsilon} \frac{\epsilon e B}{m_\epsilon^2}
= 88.6\ \epsilon\ \frac{\omega}{m_\epsilon}\
\left( \frac{\rm eV}{m_\epsilon}\right)^2
\left( \frac{B}{\rm T}\right)
\,.
\end{equation}
The above expression has been derived in leading order in an expansion
for high frequency
{\cite{Toll:1952,Klepikov:1954,Erber:1966vv,Baier:1967,Klein:1968,Heinzl:2006pn}},
\begin{equation}
\label{semiclhf}
\frac{\omega}{2m_\epsilon}\gg  1,
\end{equation}
and of high number of allowed Landau levels of the millicharged
particles \cite{Daugherty:1984tr},
\begin{eqnarray}
\nonumber
\Delta N_{\rm{p}}\!\!&=&\!\!\frac{\Delta N_{\rm{Landau}}}{2}=
\frac{1}{12}\left(\frac{\omega^{2}}{\epsilon\,eB}\right)^{2}
\left( \frac{\Delta\omega}{\omega}+ \frac{\Delta B}{2 B}\right) \gg 1
\\[1.5ex]\label{peaks}
\Leftrightarrow \epsilon\!\! &\ll &\!\! 4.9\times 10^{-3} \left(\frac{\omega}{\rm{eV}}\right)^{2}
\left(\frac{\rm{T}}{B}\right)
\left(\frac{\Delta\omega}{\omega} + \frac{\Delta B}{2 B}\right)^{\frac{1}{2}}.
\end{eqnarray}
{In the above-mentioned laser polarization experiments, the
  variation} $\Delta\omega/\omega$ is typically small compared to
$\Delta B/B\gtrsim 10^{-4}$.

Virtual production can occur even below threshold,
$\omega<2m_{\epsilon}$. Therefore, we consider both high and low
frequencies.  As long as Eq.~\eqref{peaks} is satisfied, one
has~\cite{Tsai:1975iz}
\begin{equation}\label{refraction}
n_{\parallel,\perp}^{\rm Dsp}=
1-\frac{\epsilon^{2}\alpha}{4\pi}\left(\frac{\epsilon\,eB}{m^{2}_{\epsilon}}\right)^{2}
I_{\parallel,\perp}^{\rm Dsp}(\chi),
\end{equation}
with
%
\begin{multline}
I_{\parallel,\perp}^{\rm Dsp}(\chi)\!\!=\!\!2^{\frac{1}{3}}\left(\frac{3}{\chi}\right)^{\frac{4}{3}}
\int^{1}_{0} {\rm d}v\,
\frac{\left[\left(1-\frac{v^2}{3}\right)_{\parallel},
\left(\frac{1}{2}+\frac{v^2}{6}\right)_{\perp}\right]}{(1-v^{2})^{\frac{1}{3}}}
\\
\times\tilde{e}^{\prime}_{0}\left[\begin{scriptstyle}-
\left(\frac{6}{\chi}\frac{1}{1-v^2}\right)^{\frac{2}{3}}\end{scriptstyle}\right]
\label{refrac}
\end{multline}
\begin{flushright}
$\displaystyle =
\begin{cases}  -
\frac{1}{45} \left[(14)_\parallel,(8)_\perp\right] & \text{for}\,\,\chi\ll 1\text{,} \\
\frac{9}{7}\frac{\pi^{\frac{1}{2}}2^{\frac{1}{3}}
\left(\Gamma(\left(\frac{2}{3}\right)\right)^{2}}{\Gamma\left(\frac{1}{6}\right)}
\chi^{-4/3}\left[ (3)_\parallel,(2)_\perp\right]& \text{for}\,\,\chi\gg 1\text{.}
\end{cases}
$
\end{flushright}
%
Here, $\tilde{e}_{0}$ is the generalized Airy function,
\begin{equation}
\tilde{e}_{0}(t)=\int^{\infty}_{0}{\rm d}x\,\sin\left(tx-\frac{x^3}{3}\right),
\end{equation}
and $\tilde{e}^{\prime}_{0}(t)={\rm{d}}\tilde{e}_{0}(t)/{\rm{d}}t$.

\subsubsection{Spin-0 Bosons}
\label{sec:MCB}

The optical properties of a magnetized vacuum can also be influenced
by fluctuations of charged spin-0 bosons. The corresponding
interaction Lagrangian is that of scalar QED (index ``sc''),
\begin{equation}
\label{lintsc}
{\mathcal L}^{\rm sc}= -|D_\mu(\epsilon e A)
\varphi_\epsilon|^2 - m_\epsilon^2 |\varphi_\epsilon|^2, \quad D_\mu
=\partial_\mu -\I \epsilon e A_\mu,
\end{equation}
\color{black}
with $\varphi_\epsilon$ being a complex scalar field. The induced
optical properties have not been explicitly computed before in the
literature, but can be inferred straightforwardly from the
polarization tensor found in \cite{Schubert:2000yt}. As derived in
more detail in appendices \ref{appA} and \ref{appB}, the corresponding results for dichroism
and birefringence are  similar to the familiar Dirac fermion case,
\begin{equation}
\pi_{\|,\bot}^{\rm sc}\equiv \kappa_{\parallel,\perp}^{\rm sc}\ell
= \frac{1}{2}\epsilon^3 e \alpha \frac{B \ell }{m_\epsilon}\,
T_{\parallel,\perp}^{\rm sc}(\chi )
,
\end{equation}
where
%
\begin{multline}
\label{eqDB4}
T_{\parallel,\perp}^{\text{sc}} =
\frac{2\sqrt{3}}{\pi\chi}
\int\limits_0^1 {\rm d}v\
K_{2/3}\left( \frac{4}{\chi}\frac{1}{1-v^2}\right)
\\
\times
\frac{\left[ \left(\frac{1}{3}v^2\right)_\parallel,
\left(\frac{1}{2} -\frac{1}{6}v^2\right)_\perp
\right]}{(1-v^2)}
\end{multline}
\begin{flushright}
$\displaystyle=
\begin{cases}
\frac{1}{2} \sqrt{\frac{3}{2}}\ {\rm e}^{-4/\chi}\
\left[(0 )_\parallel,(\frac{1}{4})_\perp\right] & \text{for}
\,\,\chi\ll 1\,\,\text{,} \\
\frac{\pi}{\Gamma\left(\frac{1}{6}\right)\Gamma\left(\frac{13}{6}\right)}
\chi^{-1/3}\left[ (\frac{1}{6})_\parallel,(\frac{1}{2})_\perp\right]&
 \text{for} \,\,\chi\gg 1\,\,\text{.}
\end{cases}$
\end{flushright}
%
The zero coefficient in Eq.~(\ref{eqDB4}) holds, of course, only to
leading order in this calculation. We observe that the $\bot$ mode
dominates absorption in the scalar case in contrast to the spinor
case. Hence, the induced rotation of the laser probe goes into
opposite directions in the two cases, bosons and fermions.

The refractive indices induced by scalar fluctuations read
\begin{equation}
n_{\parallel,\perp}^{\rm sc}=1-\frac{\epsilon^{2}\alpha}{4\pi}
\left(\frac{\epsilon\,eB}{m^{2}_{\epsilon}}\right)^{2}
I_{\parallel,\perp}^{\rm sc}(\chi),
\end{equation}
with
%
\begin{multline}
I_{\parallel,\perp}^{\text{sc}}(\chi)
\!\!=\frac{2^{\frac{1}{3}}}{2}\left(\frac{3}{\chi}\right)^{\frac{4}{3}}
\int^{1}_{0} {\rm d}v\,
\frac{\left[\left(\frac{v^2}{3}\right)_{\parallel},
\left(\frac{1}{2}-\frac{v^2}{6}\right)_{\perp}\right]}
{(1-v^{2})^{\frac{1}{3}}}
\\[1.5ex]
{\times\tilde{e}^{\prime}_{0}
\left[
  \begin{scriptstyle}-
    \left(\frac{6}{\chi}\frac{1}{1-v^2}\right)^{\frac{2}{3}}
  \end{scriptstyle}
\right]}
\label{refracsc}
\end{multline}
\begin{flushright}
$\displaystyle=
  \begin{cases}  -
    \frac{1}{90} \left[(1)_\parallel,(7)_\perp\right]
    & \text{for} \,\,\chi\ll 1\,\,\text{,}\\
    \frac{9}{14}\frac{\pi^{\frac{1}{2}}2^{\frac{1}{3}}
      \left(\Gamma\left(\frac{2}{3}\right)\right)^{2}}
    {\Gamma\left(\frac{1}{6}\right)}
    \chi^{-4/3}\left[
    (\frac{1}{2})_\parallel,(\frac{3}{2})_\perp\right]& \text{for}
    \,\,\chi\gg 1\,\,\text{.}
  \end{cases}$
\end{flushright}
%
Again, the polarization dependence of the refractive indices renders
the magnetized vacuum birefringent. We observe that the induced
ellipticities for the scalar and the spinor case go into opposite
directions. In particular, for small $\chi$, the $\bot$ mode is slower
for the scalar case, supporting an ellipticity signal which has the
same sign as that of Nitrogen\footnote{The sign of an
ellipticity signal can actively be checked with a residual-gas
analysis. Filling the cavity with a gas with a known classical
Cotton-Mouton effect of definite sign, this effect can interfere
constructively or destructively with the quantum effect, leading to
characteristic residual-gas pressure dependencies of the total signal
\cite{PVLASICHEP,Cantatore:IDM2006}.}. For the spinor case, it is the
other way round. As a nontrivial cross-check of our results for the
scalar case, note that the refractive indices for $\chi\ll 1$
precisely agree with the (inverse) velocities computed in
Eqs.~\eqref{eqB3} and \eqref{eqB4} from the Heisenberg-Euler effective
action of scalar QED.

We conclude that a careful determination of the signs of ellipticity
and rotation in the case of a positive signal can distinguish between
spinor and scalar fluctuating particles.\footnote{{In the sense of
classical optics, the ellipticities of the various scenarios discussed
here are indeed associated with a definite and unambiguous sign. This
is not the case for the sign of the rotation which also depends on the
experimental set up: in all our scenarios, the polarization axis is
rotated towards the mode with the smallest probability exponent $\pi$
in Eq.~\eqref{delthet}. In the sense of classical optics, this can be
either sign depending on the initial photon polarization relative to
the magnetic field. In this work, the notion of the sign of rotation
therefore refers to the two experimentally distinguishable cases of
either $\pi_\|>\pi_\bot$ or $\pi_\|<\pi_\bot$.  }}

Finally, let us briefly comment on the case of {having both
  fermions and bosons.  If there is} an identical number of bosonic
and fermionic degrees of freedom with exactly the same masses and
millicharges, i.e. if the millicharged particles appear in a
supersymmetric fashion in complete supersymmetric chiral multiplets,
one can check that the signals cancel. An exactly supersymmetric set
of millicharged particles would cause neither an ellipticity signal
nor a rotation of the polarization and one would have to rely on other
detection principles as, for example, Schwinger pair production in
accelerator cavities \cite{Gies:2006hv}.  However, in nature
supersymmetry is broken resulting in different masses for bosons and
fermions.  Now, the signal typically decreases rather rapidly for
large masses (more precisely when $\chi\sim 1/m^{3}_{\epsilon}$
becomes smaller than one) and the lighter particle species will give a
much bigger contribution. Accordingly, for a sufficiently large mass
splitting the signal would look more or less as if we had only the
lighter particle species, be it a fermion or a boson.

\begin{figure*}[ht!]
{\centering\includegraphics[width=0.9\linewidth]{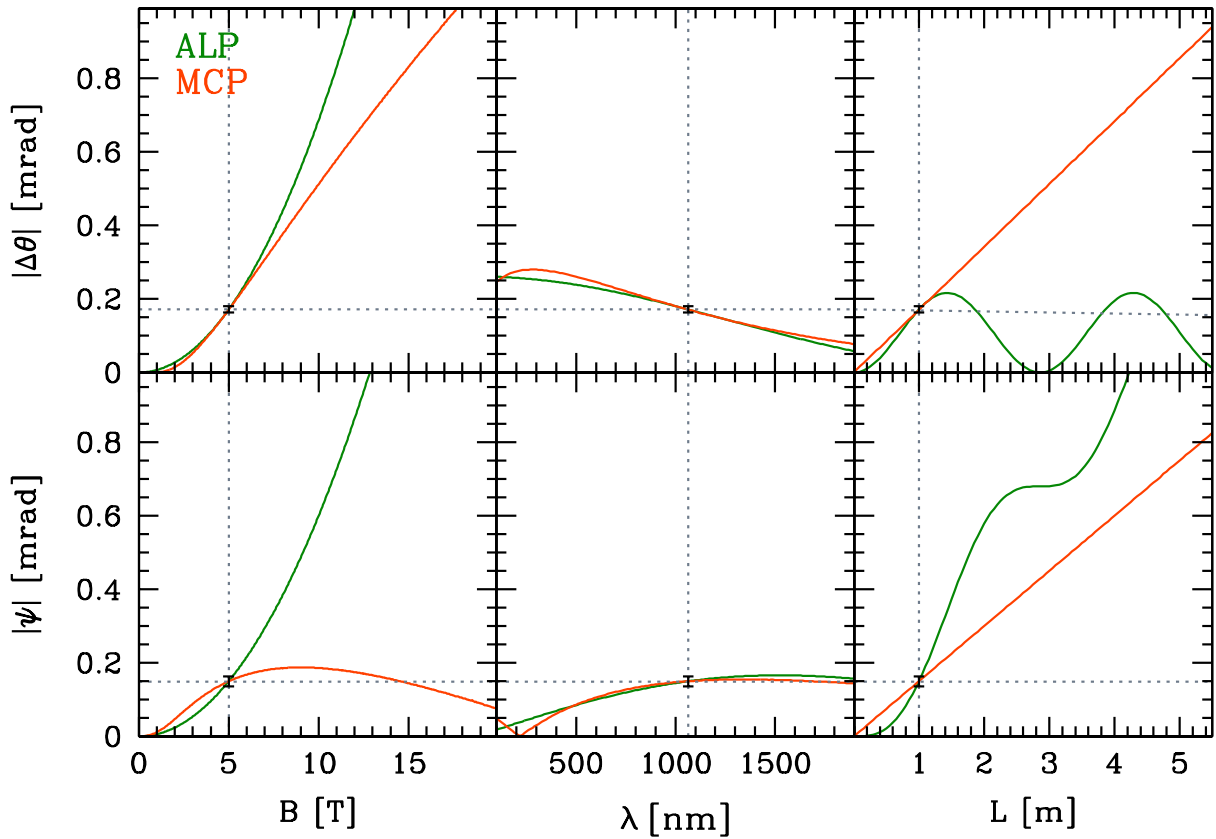}}
\caption{Dependence of the rotation and ellipticity signals on the
strength of the magnetic field $B$, the wavelength $\lambda$ of the
laser, and the length $L$ of the magnetic region inside the
cavity. For ALPs (dark green) and MCPs (light red). The crossing of
the blue dotted lines corresponds to the PVLAS published rotation and
preliminary ellipticity signal for $B=5$~T, $\lambda=1064$~nm, and
$L=1$~m.}
\label{dependencies}
\end{figure*}

\section{Distinguishing between different Scenarios}\label{sec3}

In principle, one can set up a series of different experiments
distinguishing between the different scenarios, ALPs or MCPs. For
example, a positive signal in a light-shining-through-wall experiment
\cite{Sikivie:1983ip,Anselm:1986gz,Gasperini:1987da,VanBibber:1987rq,%
Ruoso:1992nx,Ringwald:2003ns,Gastaldi:2006fh,Pugnat:2005nk,Rabadan:2005dm,%
Cantatore:Patras,Kotz:2006bw,Baker:Patras,Rizzo:Patras,ALPS} would be
a clear signal for the ALP interpretation, whereas detection of a dark
current that is able to pass through walls would be a clear signal for
the MCP hypothesis \cite{Gies:2006hv}. But even with a PVLAS-type
experiment that measures only the rotation and ellipticity signals,
one can collect strong evidence favoring one and disfavoring other
scenarios.

\begin{table}[b]
\renewcommand{\arraystretch}{1.3}
\begin{tabular}{c||c|c}
 &\begin{minipage}[c][0.6cm][c]{3cm} $n_\|> n_\bot$ \end{minipage}&
\begin{minipage}[c][0.6cm][c]{3cm} $ n_\|< n_\bot$ \end{minipage}\\
\hhline{=::=|=}
$\pi_\|> \pi_\bot$ & \begin{minipage}[c][1.2cm][c]{3cm}
  ALP 0${}^{-}$ or\\MCP $\frac{1}{2}$ (small $\chi$)
  \end{minipage}&   MCP $\frac{1}{2}$ (large $\chi$)\\
\hline
$\pi_\|< \pi_\bot$ &MCP $0$ (large $\chi$) & \begin{minipage}[c][1.2cm][c]{3cm}
  ALP 0${}^{+}$ or\\MCP $0$ (small $\chi$)
  \end{minipage}
\end{tabular}
\caption{Summary of the allowed particle-physics interpretation
  arising from a sign analysis of birefringence induced by
  different refractive indices $n_{\|,\bot}$ and dichroism induced by
  different probability exponents $\pi_{\|,\bot}$.
\label{tab1}
}
\end{table}

Performing one measurement of the absolute values of rotation and
ellipticity, one can typically find values for the masses and
couplings in all scenarios, such that the predicted rotation and
ellipticity is in agreement with the experiment.

One clear distinction can already be made by measuring the sign of the
ellipticity and rotation signals.  In the ALP scenario, a measurement
of the sign of either the rotation or the ellipticity is sufficient to
decide between a scalar or pseudoscalar. Measuring the sign of both
signals already is a consistency check; {if the signal signs turn out
to be inconsistent, the ALP scenarios for both the scalar and the
pseudoscalar would be ruled out}.  In the MCP scenario, a measurement
of the sign of rotation decides between scalars and
fermions. If only the sign of the ellipticity signal is
measured, both options still remain, since the sign of the
ellipticity changes when one moves from large to small masses: the
hierarchy of the refractive indices is inverted in the region of
anomalous dispersion. But at least the sign tells us if we are in the
region of large or small masses, {corresponding to a small or large
$\chi$ parameter, cf. Eq.~\eqref{chi}. This sign analysis is
summarized in Table \ref{tab1}.}

More information can be obtained by varying the parameters of the
experiment. In principle, we can vary all experimental parameters
appearing in Eqs.~\eqref{dicALP}, \eqref{birALP}, \eqref{absorption}
and \eqref{refraction}: the strength of the magnetic field $B$, the
frequency of the laser $\omega$, and the length of the magnetic field
inside the cavity ${L}$.

Let us start with the magnetic field dependence. For the ALP scenario
both rotation and ellipticity signals are proportional to $B^2$,
\begin{equation}
\Delta\theta^\text{ALP} \sim B^{2}, \quad \psi^\text{ALP}\sim B^{2}
\end{equation}
whereas for MCP's we have
\begin{eqnarray}
 \Delta\theta^\text{MCP}\!\!&\sim&\!\!\bigg\{
\begin{array}{ll}
\exp\left(-\frac{\text{const}}{B}\right) &  \quad B\,\, \text{small}\\
B^{\frac{2}{3}} & \quad B\,\, \text{large}
\end{array}
\\\nonumber
\psi^\text{MCP}\!\!&\sim&\!\! 
\bigg\{
\begin{array}{ll}
B^2 &  \quad B\,\, \text{small}\\
B^{\frac{2}{3}}&  \quad B\,\, \text{large}.\\
\end{array}
\end{eqnarray}
In the left panels of Fig.~\ref{dependencies} we demonstrate the
different behavior (for the ellipticity signal the
$B^{\frac{2}{3}}$-dependence is not yet visible as it appears only at
much stronger fields). The model parameters for ALPs and
MCPs are chosen such that the absolute value of $\Delta\theta$ and
$\psi$ matches the PVLAS results ($\lambda=1064$~nm, $B=5$~T, and
$L=1$~m) shown as the crossing of the dotted lines together with their
statistical errors.  In a similar manner, the signals also depend on
the wavelength of the laser light, which is shown in the center panels
of Fig.~\ref{dependencies}.

Finally, there is one more crucial difference between the ALP and the
MCP scenario. Production of a single particle can occur coherently.
This leads to a faster growth of the signal
\begin{equation}
 \Delta\theta^\text{ALP}\sim {L}^{2},\quad
 \psi^\text{ALP}\sim{L}^{2} \quad\,\, {L}\,\,\text{small}.
\end{equation}
In the MCP scenario, however, the produced particles are essentially
lost and we have only a linear dependence on the length of the
interaction region,
\begin{equation}
 \Delta\theta^\text{MCP}\sim{L},\quad\psi^\text{MCP}\sim{L}.
\end{equation}
This is shown in the right panels of Fig.~\ref{dependencies}. 

We conclude that studying the dependence of the signal on the
parameters of the experiment can give crucial information to decide
between the ALP and MCP scenarios, as we will also see in the following
section.

\begin{table}
\renewcommand{\arraystretch}{1.3}
\begin{center}
\begin{tabular}{ccc}
\hline\hline
\multicolumn{3}{c}{\makebox[8cm][c]{\bf BFRT experiment}}\\
\hline
\multicolumn{3}{c}{{\bf Rotation}\hfill($L=8.8$~m, $\lambda=514.5$~nm, $\theta=\nicefrac{\pi}{4}$)}\\[0.1cm]
\makebox[2.5cm][c]{$N_\text{pass}$}&\makebox[2.5cm][c]{$\left|{\Delta\theta}\right|\,[\text{nrad}]$}&\makebox[2.5cm][c]{${\Delta\theta}_\text{noise}\,[\text{nrad}]$}\\[0.1cm]
$254$&$0.35$&$0.30$\\
$34$&$0.26$&$0.11$\\
\hline
\multicolumn{3}{c}{{\bf Ellipticity}\hfill($L=8.8$~m, $\lambda=514.5$~nm, $\theta=\nicefrac{\pi}{4}$)}\\[0.1cm]
\makebox[2.5cm][c]{$N_\text{pass}$}&\makebox[2.5cm][c]{$\left|\psi\right|\,[\text{nrad}]$}&\makebox[2.5cm][c]{$\psi_\text{noise}\,[\text{nrad}]$}\\[0.1cm]
$578$&$40.0$&$11.0$\\
$34$&$1.60$&$0.44$\\
\hline
\multicolumn{3}{c}{{\bf Regeneration}\hfill($L=4.4$~m, $\langle\lambda\rangle=500$~nm,  $N_\text{pass}=200$)}\\[0.1cm]
\makebox[2.5cm][c]{$\theta\, [\text{rad}]$}&\multicolumn{2}{c}{\makebox[5cm][c]{rate $[\text{Hz}]$}}\\[0.1cm]
$0$&\multicolumn{2}{c}{$-0.012\pm0.009$}\\
$\nicefrac{\pi}{2}$&\multicolumn{2}{c}{$0.013\pm0.007$}\\
\hline\hline
\end{tabular}

\end{center}
\caption{The vacuum rotation ${\Delta\theta}$, ellipticity $\psi$ and
photon regeneration rate from the BFRT~\cite{Cameron:1993mr}
experiment. For simplicity we take the noise level
${\Delta\theta}_\text{noise}$ and $\psi_\text{noise}$ quoted in
Ref.~\cite{Cameron:1993mr} as the standard deviation
$\sigma_{\Delta\theta}$ and $\sigma_\psi$.  For the polarization data,
BFRT used a magnetic field with time-varying amplitude $B=B_0+\Delta
B\cos(\omega_m t+\phi_m)$, where $B_0=3.25$~T and $\Delta B=0.62$~T
(cf. Appendix \ref{appC}). For photon regeneration, they employed
$B=3.7$~T. }
\label{BFRTresults}
\end{table}

\section{Confrontation with Data}\label{sec4}

In this Section, we want to confront the prediction of the ALP and MCP
scenarios for vacuum magnetic dichroism, birefringence, and photon
regeneration with the corresponding data from the
BFRT~\cite{Cameron:1993mr} and
PVLAS~\cite{Zavattini:2005tm,PVLASICHEP,Cantatore:IDM2006}
collaborations, as well as from the Q\&A experiment
\cite{Chen:2006cd}. The corresponding experimental findings are
summarized in Tables~\ref{BFRTresults}, \ref{PVLASresults}, and
\ref{QandAresults}, respectively.

\begin{table}
\renewcommand{\arraystretch}{1.3}
\begin{center}
\begin{tabular}{cc}
\hline\hline
\multicolumn{2}{c}{\makebox[8cm][c]{\bf PVLAS experiment}}\\
\hline
\multicolumn{2}{c}{{\bf Rotation}\hfill($L=1$~m, $N_\text{pass}=44000$, $\theta=\nicefrac{\pi}{4}$)}\\[0.2cm]
 \makebox[2.5cm][c]{$\lambda\,\, [\text{nm}]$}&\makebox[5cm][c]{$\left|{\Delta\theta}\right|\,[10^{-12}\,\text{rad}/\text{pass}]$}\\[0.2cm]
 $1064$&$3.9\pm0.2$\\
$532$&$6.3\pm1.0$ {\bf (preliminary)}\\
\hline
\multicolumn{2}{c}{{\bf Ellipticity}\hfill($L=1$~m, $N_\text{pass}=44000$, $\theta=\nicefrac{\pi}{4}$)}\\[0.2cm]
\makebox[2.5cm][c]{$\lambda\,\, [\text{nm}]$}&\makebox[5cm][c]{$\psi\,[10^{-12}\,\text{rad}/\text{pass}]$}\\[0.2cm]
$1064$&$-3.4\pm0.3$ {\bf (preliminary)}\\
$532$&$-6.0\pm0.6$ {\bf (preliminary)}\\
\hline\hline
\end{tabular}

\end{center}
\caption{The vacuum rotation ${\Delta\theta}$ and ellipticity $\psi$
  per pass measured by 
PVLAS, for $B=5$~T. 
The rotation of polarized laser light
  with $\lambda=1064$~nm is published in Ref.~\cite{Zavattini:2005tm}.
  Preliminary results are taken from
  Refs.~\cite{PVLASICHEP,Cantatore:IDM2006} {and are used here
  for illustrative purposes only}. }
\label{PVLASresults}
\end{table}

\begin{table}
\renewcommand{\arraystretch}{1.3}
\begin{center}
\begin{tabular}{ccc}
\hline\hline
\multicolumn{3}{c}{\makebox[8cm][c]{\bf Q\&A experiment}}\\
\hline
\multicolumn{3}{c}{{\bf Rotation}\hfill($L=1$~m, $\lambda=1064$~nm, $\theta=\nicefrac{\pi}{4}$)}\\[0.1cm]
\makebox[2.5cm][c]{$N_\text{pass}$}&\multicolumn{2}{c}{\makebox[5cm][c]{$\Delta\theta\,[\text{nrad}]$}}\\[0.1cm]
$18700$&\multicolumn{2}{c}{$-0.4\pm5.3$}\\
 \hline\hline
\end{tabular}

\end{center}
\caption{The vacuum rotation ${\Delta\theta}$ from the Q\&A
experiment~\cite{Chen:2006cd} experiment
($B=2.3$~T).
}
\label{QandAresults}
\end{table}

In the following we combine these results in a simple statistical
analysis. For simplicity, we assume that the likelihood function $L_i$
of the rotation, the ellipticity and the photon regeneration rate
follows a Gaussian distribution in each measurement $i$ with mean
value and standard deviation as indicated in
Tables~\ref{BFRTresults}-\ref{QandAresults}.  In the case of the BFRT
upper limits, we approximate the likelihood functions by\footnote{We
set the negative photon regeneration rate (Tab.~\ref{BFRTresults}) at
BFRT for $\theta=0$ equal to zero.}
$L\propto\exp((\psi-\psi_\text{hypo})^2/(2\psi_\text{noise}^2))$.
Taking these inputs as statistically independent values we can
estimate the combined log-likelihood function as $\ln L \approx \sum_i
\ln L_i$ \cite{Yao:2006px}.  With these assumptions the method of
maximum likelihood is equivalent to the method of least squares with
$\chi^2=\text{const} - 2\sum_i\ln L_i$. A more sophisticated
statistical analysis is beyond the scope of this work and requires
detailed knowledge of the data analysis.

\begin{figure*}[ht!]
\centering
\includegraphics[width=0.48\linewidth]{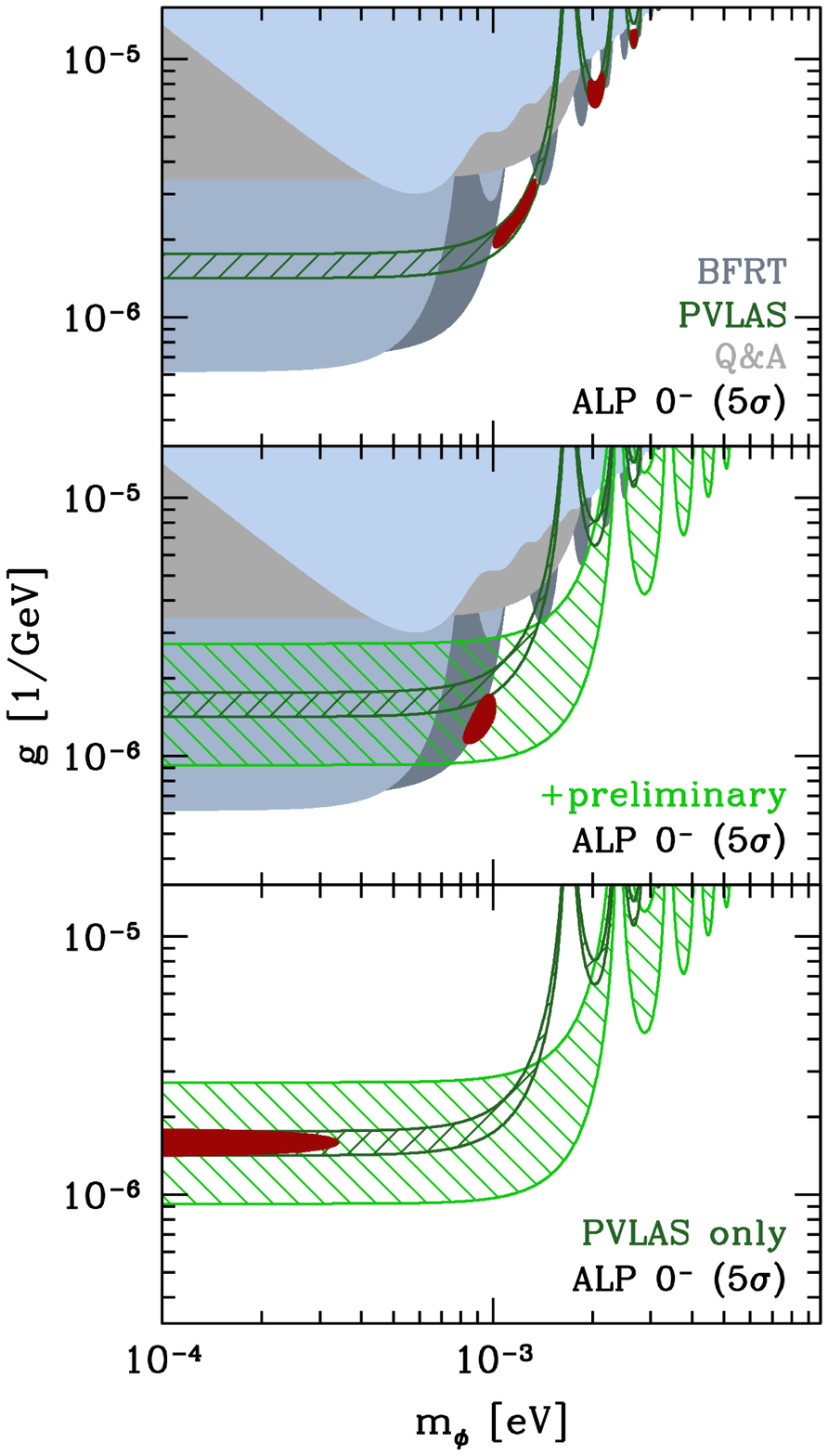}\hfill
\includegraphics[width=0.48\linewidth]{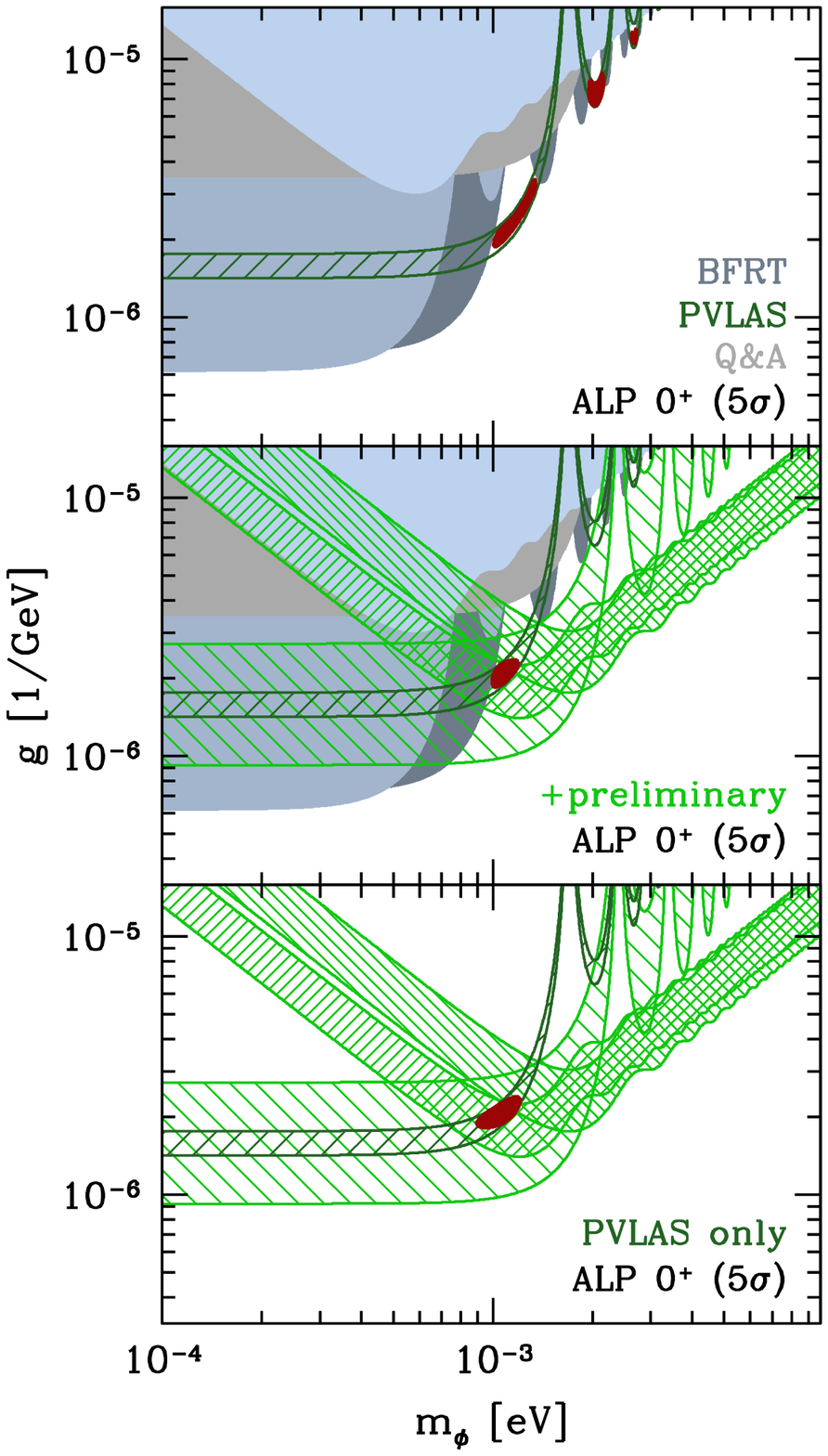}
\caption{ALP: The $5\sigma$ confidence level of the model
  parameters (red). The blue shaded regions arise from the BFRT upper limits
  for regeneration (darkblue), rotation (blue) and ellipticity
  (lightblue).  The gray shaded region is the Q\&A upper limit
  for rotation.  The bands show the PVLAS $5\sigma$ C.L.s for rotation
  (coarse-hatched) and ellipticity (fine-hatched) with
  $\lambda=532$~nm (left-hatched) and $\lambda=1064$~nm
  (right-hatched), respectively.  The darkgreen band shows the
  published result for rotation with $\lambda=1064$~nm.  {The
  lightgreen bands result from an inclusion of preliminary data from
  PVLAS. The upper panels show the fit to the published data; the
  center panels include also the preliminary data from PVLAS, and the
  lower panels depict the fit using only PVLAS data. The preliminary
  data is only used to demonstrate the potential to distinguish
  between the different scenarios.}}
 \label{ALPfit}

\end{figure*}

\begin{figure*}[ht!]
\centering
\includegraphics[width=0.48\linewidth]{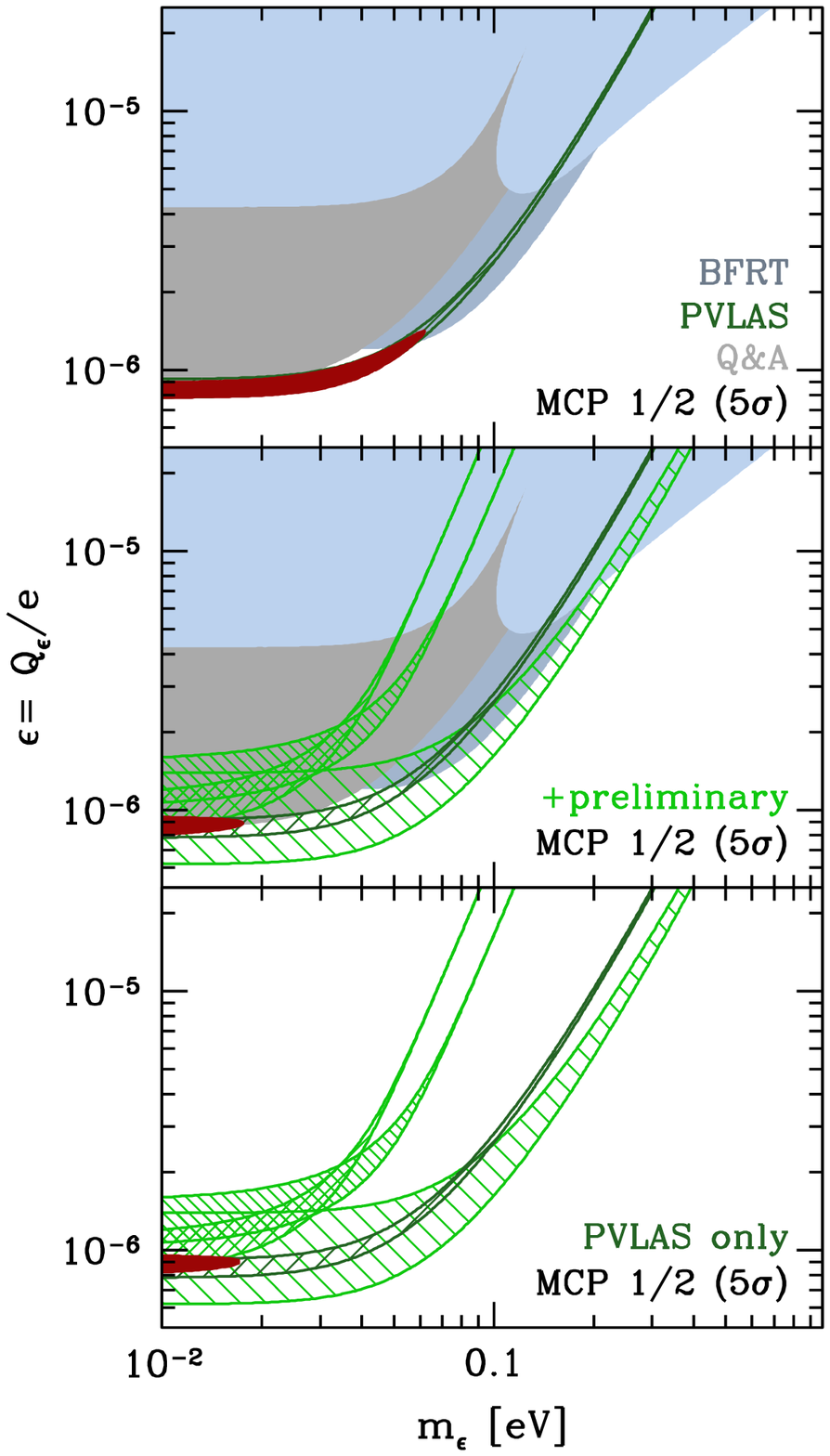}\hfill
\includegraphics[width=0.48\linewidth]{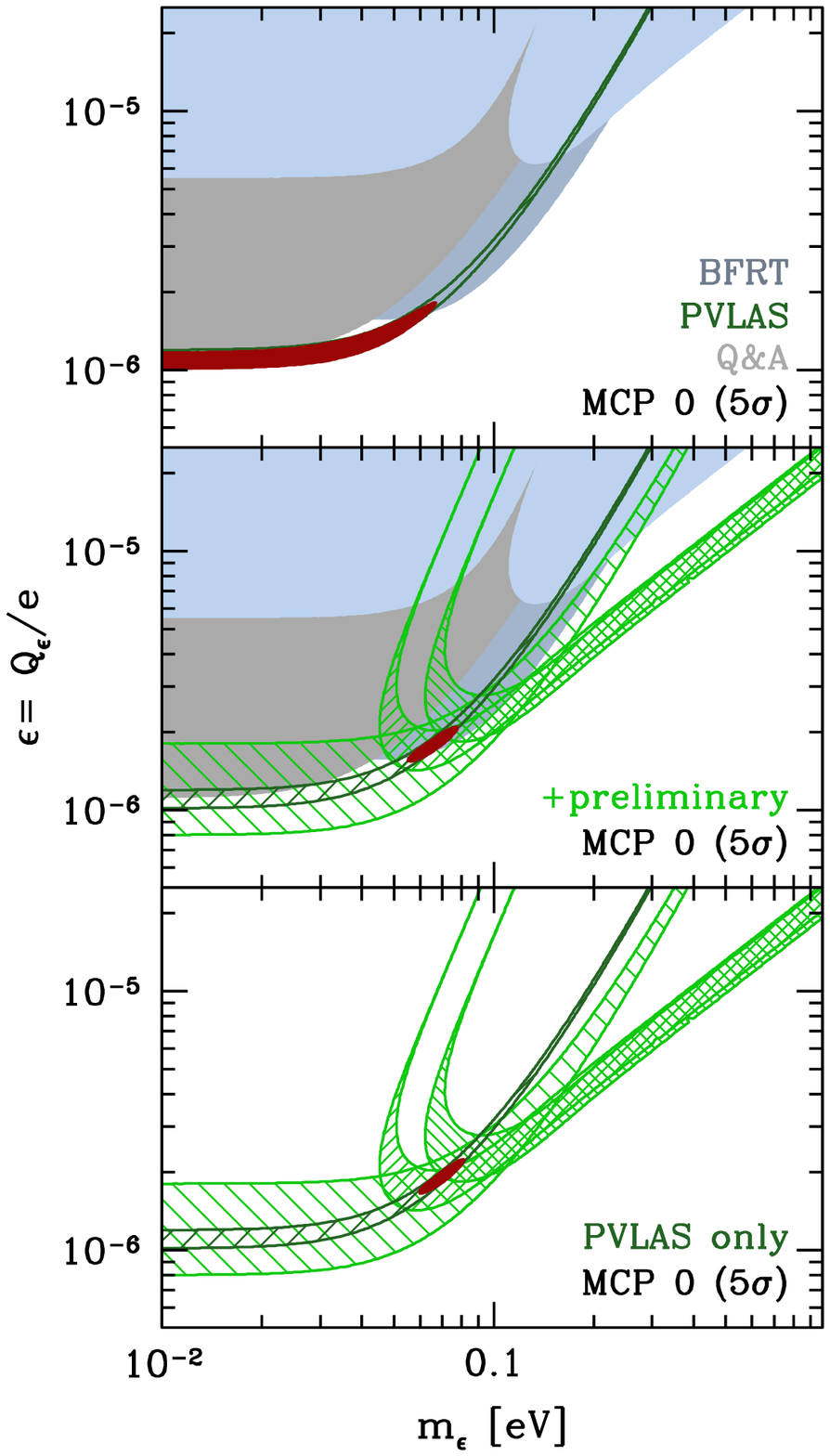}
\caption{MCP: The $5\sigma$ confidence level of the model parameters
  (red). The blue shaded regions arise from the BFRT upper limits for
  rotation (blue) and ellipticity (lightblue). The gray shaded region
  is the Q\&A upper limit for rotation.  The bands show the PVLAS
  $5\sigma$ C.L.s for rotation (coarse-hatched) and ellipticity
  (fine-hatched) with $\lambda=532$~nm (left-hatched) and
  $\lambda=1064$~nm (right-hatched), respectively.  The darkgreen band
  shows the published result for rotation with $\lambda=1064$~nm. {The
  lightgreen bands result from an inclusion of preliminary data from
  PVLAS. The upper panels show the fit to the published data; the
  center panels include also the preliminary data from PVLAS, and the
  lower panels depict the fit using only PVLAS data. The preliminary
  data is only used to demonstrate the potential to distinguish
  between the different scenarios. The preliminary PVLAS value for the
  sign of the ellipticity singles out the large-$\chi$ (small-mass)
  branch of the fermionic MCP $\frac{1}{2}$ and the small-$\chi$
  (large-mass) branch of the scalar MCP 0, cf. Table \ref{tab1}, as is
  visible in the center and lower panels.} }
\label{MCPfit}
\end{figure*}

\subsection{ALP hypothesis}

\begin{table}
\renewcommand{\arraystretch}{1.3}
\begin{tabular}{c||c|c|c|c}
\begin{minipage}[c][0.6cm][c]{3cm} $\chi^2/$d.o.f.
\end{minipage} & ALP 0${}^{-}$ & ALP 0${}^{+}$ & MCP $\frac{1}{2}$   &
 MCP 0   \\
\hhline{=::=|=|=|=}  
\begin{minipage}[c][1.2cm][c]{3cm}\centering BFRT, PVLAS, Q\&A
   published data\\ (d.o.f.$=6$)
\end{minipage}&  1.3 & 0.8 &  7.4 & 7.3   \\
\hline
\begin{minipage}[c][1.2cm][c]{3cm}\centering + PVLAS \\ 
preliminary data\\ (d.o.f.$=9$)
\end{minipage} &  62.0 &  6.3  &  15.7 &  12.0 \\
\hline
\begin{minipage}[c][1.2cm][c]{3cm}\centering  only PVLAS \\ 
pub.~+ prelim.~data\\ (d.o.f.$=2$)
\end{minipage} 
  & 118.4 &    18.9  & 40.0 &  15.7 
\end{tabular}
\caption{Summary of the $\chi^2$/d.o.f.~analysis for the different
  scenarios and based on different data sets. Rows and columns
  correspond to the rows and columns of panels in Figs. \ref{ALPfit}
  and \ref{MCPfit}. 
\label{tab2}
}
\end{table}

Figure~\ref{ALPfit} shows the results of a fit based on the
pseudoscalar (left panels) or scalar (right panels) ALP hypothesis.
The BFRT upper limits\footnote{As far as photon regeneration at BFRT
is concerned, their photon detection efficiency $\eta$ was
approximately 5.5\%.  Their laser spectrum with average power $\langle
P\rangle \approx3$~W and average photon flux $\dot{N}_0=\langle
P\rangle/\omega$ was dominated by the spectral lines $488$~nm and
$514.5$~nm. We took an average value of $500$~nm in our fitting
procedure.}  are shown by blue-shaded regions. The Q\&A upper rotation
limit is depicted as a gray-shaded region, but this limit exerts
little influence on the global fit in the ALP scenario. The PVLAS
results are displayed as green bands according to the $5\sigma$
confidence level (C.L.) with dark green corresponding to published
data and light green corresponding to preliminary results. The
resulting allowed parameter regions at 5$\sigma$ CL are depicted as
red-filled islands or bands.

Both upper panels show the result from all published data of all three
experiments. Here, the results for scalar or pseudoscalar ALPs are
very similar: in addition to the allowed 5$\sigma$ region at
$m_\phi\simeq 1\dots 2\times 10^{-3}$~eV also reported by PVLAS
\cite{Zavattini:2005tm}, we observe further allowed islands for larger
mass values. The $\chi^2$/d.o.f.~(degrees of freedom) values for the fits are both
acceptable with a slight preference for the scalar ALP
($\chi^2$/d.o.f.=0.8) in comparison with the pseudoscalar ALP
($\chi^2$/d.o.f.=1.3), cf. Table \ref{tab2}. 

This degeneracy between the scalar and the pseudo-scalar ALP scenario
is lifted upon the inclusion of the preliminary PVLAS data (center
panels), since the negative sign of the birefringence signal with
$n_\|<n_\bot$ strongly prefers the scalar ALP scenario. In addition,
the size of the preliminary ellipticity result is such that the higher
mass islands are ruled out, and the low mass island settles around
$m_{\phi}\simeq 10^{-3}$~eV and $g\simeq 2\times 10^{-6}$~GeV${}^{-1}$.
The results from a fit to PVLAS data only (published and preliminary)
as displayed in the lower panels of Fig.~\ref{ALPfit} remain similar.

\subsection{MCP hypothesis}

Figure~\ref{MCPfit} shows the results of a fit based on the fermionic
 (left panels) or scalar (right panels) MCP hypothesis.  The MCP
 hypothesis gives similar results for scalars and fermions if only the
 published data is included in the fit (upper panels). MCP masses
 $m_{\epsilon}$ larger than 0.1~eV are ruled out by the upper limits
 of BFRT. But the 5$\sigma$ CL region shows a degeneracy towards
 smaller masses. It is interesting to observe that the available Q\&A
 data already approaches the ballpark of the PVLAS rotation signal in
 the light of the MCP hypothesis, whereas it is much less relevant for
 the ALP hypothesis.
  
  Including the PVLAS preliminary data, the fit for fermionic MCPs
  becomes different from the scalar MCP case: because of the negative
  sign of the birefringence signal, only the
  large-$\chi$/small-$m_{\epsilon}$ branch remains acceptable for the
  fermionic MCP, whereas the small-$\chi$/large-$m_\epsilon$ branch is
  preferred by the scalar MCP, cf. Table \ref{tab1}. A $\chi^2/$d.o.f.~comparison 
  between the fermionic MCP ($\chi^2/$d.o.f.$=15.7$) and the
  scalar MCP ($\chi^2/$d.o.f.$=12.0$) points to a slight preference for
  the scalar MCP scenario. 

  This preference is much more pronounced in the fit to the PVLAS data
  (published + preliminary) only, cf. Table \ref{tab2}. The best MCP
  candidate would therefore be a scalar particle with mass
  $m_\epsilon\simeq0.07$~eV and charge parameter $\epsilon\simeq 2\times 
  10^{-6}$.  

\subsection{ALP vs. MCP}

Let us first stress that the partly preliminary status of the data
used for our analysis does not yet allow for a clear preference of
either of the two scenarios, ALP or MCP. Based on the published data
only, the ALP scenarios give a better fit, since the upper limits by
BFRT and Q\&A leave an unconstrained parameter space open to the PVLAS
rotation data. By contrast, the BFRT and Q\&A upper limits already
begin to restrict the MCP parameter space of the PVLAS rotation signal
in a sizable manner, which explains the better $\chi^2$/d.o.f.~for the
ALP scenario. 

Based on the (in part preliminary) PVLAS data alone, the MCP scenario
would be slightly preferred in comparison with the ALP scenario, see
Table \ref{tab2}, bottom row. The reason is that the PVLAS
measurements of birefringence and rotation for the different laser
wavelengths show a better internal compatibility in the scalar MCP
case than in the scalar ALP scenario.

\section{Conclusions}\label{conclusions}

The signal observed by PVLAS -- a rotation of linearly polarized laser
light induced by a transverse magnetic field -- has generated a great
deal of interest over the recent months. Since the signal has found no
explanation within standard QED or from other standard-model sectors,
it could be the first direct evidence of physics beyond the standard
model.

The proposed attempts to explain this result fall into two categories:
\\
1. conversion of laser photons into a single neutral spin-0 particle
(scalar or pseudoscalar) coupled to two photons (called axion-like
particle or ALP) and\\
2. pair production of fermions or bosons with a small electric charge
(millicharged particles or MCPs).\\
The corresponding actions associated with these two proposals
  should be viewed as pure low-energy effective field theories which
  are valid at laboratory scales at which the experiments operate. A
  naive extrapolation of these theories to higher scales generically
  becomes incompatible with astrophysical bounds.  In this paper, we
have compared the different low-energy effective theories in
light of the presently available data from optical experiments.

We have summarized the formulas for rotation and ellipticity in the
different scenarios and contributed new results for millicharged
scalars.  We have then studied how optical experiments can provide for
decisive information to discriminate between the different scenarios:
this information can be obtained in the form of size and sign of
rotation and ellipticity and their dependence on experimental
parameters like the strength of the magnetic field, the wavelength of
the laser and the length of the magnetic region.

Our main results are depicted in Figs. \ref{ALPfit} and \ref{MCPfit}
which show the allowed parameter regions for the different scenarios.
On the basis of the published data, none of the scenarios can
  currently be excluded. The remaining open parameter regions should be
  regarded as good candidates for the target regions of future
  experiments. As the preliminary PVLAS data illustrates, near future
optical measurements can further constrain the parameter space and
even decide between the different scenarios. For instance, a
  negative ellipticity $n_\|<n_\bot$ together with a rotation
  corresponding to probability exponents $\pi_\|>\pi_\bot$ would rule
  out the scalar or pseudo-scalar ALP interpretation altogether.

Be it from optical experiments like PVLAS or from the proposed 
``light/dark current shining through a wall'' experiments, we will soon know 
more about the particle interpretation of PVLAS.

\section{Acknowledgments}

The authors would like to thank Stephen L.~Adler, Giovanni Cantatore,
Walter Dittrich, Angela Lepidi, Axel Lindner, Eduard Masso, and
Giuseppe Ruoso for insightful discussions.  H.G.~acknowledges support
by the DFG under contract Gi 328/1-3 (Emmy-Noether program).

\appendix

\section{Birefringence in the small-$\omega$ limit: effective action
  approach}\label{appA}

{
Since the sign of the ellipticity signaling birefringence can be a
decisive piece of information, distinguishing between the spin
properties of the new hypothetical particles, let us check our results
with the effective-action approach \cite{Dittrich:2000zu}.}
Since the formulas in this appendix are equally valid for the
  MCP scenario as well as standard QED, we denote the coupling and
  mass of the fluctuating particle with $\ta$, or $\te$, and $\tm$
  with the dictionary:
  \begin{eqnarray}
    \text{MCP:}&&\quad \te=\epsilon e,\quad \ta=\epsilon^2
    \alpha,\quad \tm=m_\epsilon, \nonumber\\
    \text{QED:}&&\quad \te= e,\quad \ta=
    \alpha,\quad \tm=m_e. \label{preEq1}
  \end{eqnarray}

The effective action in one-loop approximation can be written as
\begin{equation}
\Gamma[A]=S_{\text{cl}}[A]+\Gamma^1[A]
= -\int_x \mathcal F + \Gamma^1[A], \label{Eq1}
\end{equation}
where we have introduced the field-strength invariant $\mathcal F$
corresponding to the Maxwell action. The two possible invariants are
\begin{equation}
\mathcal F = \frac{1}{4} F_{\mu\nu} F^{\mu\nu}=\frac{1}{2}
(\vec{B}^2 - \vec{E}^2), \quad
\mathcal G = \frac{1}{4} F_{\mu\nu} \Fd^{\mu\nu} = -\vec{E}\cdot
  \vec{B}. \label{eq2}
\end{equation}
with $\Fd_{\mu\nu}=\frac{1}{2} \epsilon_{\mu\nu\kappa\lambda}
F^{\kappa\lambda}$. Also useful are the two secular invariants $a,b$,
corresponding to the eigenvalues of the field strength tensor,
\begin{equation}
a=\sqrt{\sqrt{\mathcal{F}^2+\mathcal{G}^2}+\mathcal{F}},\quad
b=\sqrt{\sqrt{\mathcal{F}^2+\mathcal{G}^2}-\mathcal{F}}, \label{eq3}
\end{equation}
with the inverse relations
\begin{equation}
|\mathcal{G}|=ab, \quad \mathcal{F}=\frac{1}{2} (a^2-b^2).\label{eq4}
\end{equation}
Let us start with the fermion-induced effective action, i.e., the
classic Heisenberg-Euler effective action. The one-loop contribution
reads
%
\begin{multline}
\Gamma^1_{\text{Dsp}}
= \frac{1}{8\pi^2} \int_x \int_0^\infty \frac{ds}{s^3}\, \E^{-\I \tm^2 s}\\
\times\!\left( \te as \cot (\te as)\, \te bs \coth(\te bs)
+ \frac{2}{3} (\te s)^2\mathcal{F}  -1 \right).
\end{multline}
%
Expanding this action to quartic order in the field strength results
in
\begin{equation}
\Gamma^1_{\text{Dsp}}
= \int_x \left( c^{\text{Dsp}}_{\bot}\,
  \mathcal{F}^2 +c^{\text{Dsp}}_\|
  \mathcal{G}^2 \right), \label{eq5}
\end{equation}
where the constant prefactors read
\begin{equation}
c^{\text{Dsp}}_\bot=\frac{8}{45} \frac{\ta^2}{\tm^4}, \quad
c^{\text{Dsp}}_\|=\frac{14}{45} \frac{\ta^2}{\tm^4}. \label{eq6}
\end{equation}

It is straightforward to derive the modified Maxwell equations from
\Eqref{eq5}. From these, the dispersion relations for the two
polarization eigenmodes of a plane-wave field in an external magnetic
field can be determined \cite{Dittrich:2000zu}, yielding the phase
velocities in the low-frequency limit,
\begin{equation}
v_\bot =1 -c^{\text{Dsp}}_\bot B^2 \sin^2
\theta_B,\quad
v_\| =1-c^{\text{Dsp}}_\| B^2 \sin^2
\theta_B.\label{eq7}
\end{equation}
Obviously, the $\bot$ mode is slightly faster than the $\|$ mode,
since the coefficient 
$c^{\text{Dsp}}_\bot <c^{\text{Dsp}}_\|$.

Next we turn to the effective action which is induced by charged
scalar fluctuations, i.e., the Heisenberg-Euler effective action for
scalar QED.  The one-loop contribution now reads
%
\begin{multline}
\Gamma^1_{\text{sc}}
=- \frac{1}{16\pi^2} \int_x \int_0^\infty \frac{ds}{s^3}\, \E^{-\I
  \tm^2 s}\\
\times\!\left(\frac{ \te as}{ \sin (\te as)}\,\frac{ \te bs}{
  \sinh(\te bs)} - \frac{1}{3}
  (\te s)^2\mathcal{F}  -1 \right).\label{eqB1}
\end{multline}
%
There are three differences to the fermion-induced action: the minus
sign arises from Grassmann integration in the fermionic case. The
factor of 1/2 comes from the difference between a trace over a complex
scalar and that over a Dirac spinor. The replacement of $\cot$ and
$\coth$ by inverse $\sin$ and $\sinh$ is due to the Pauli spin-field
coupling in the fermionic case.  

Expanding the scalar-induced action
to quartic order in the field strength results in
\begin{equation}
\Gamma^1_{\text{sc}} = \int_x \left( c^{\text{sc}}_{\bot}\,
  \mathcal{F}^2 +c^{\text{sc}}_\| \mathcal{G}^2 \right), \label{eqB2}
\end{equation}
where the constant prefactors this time read
\begin{equation}
c^{\text{sc}}_\bot=\frac{7}{90} \frac{\ta^2}{\tm^4}, \quad
c^{\text{sc}}_\|=\frac{1}{90} \frac{\ta^2}{\tm^4}. \label{eqB3}
\end{equation}
The velocities of the two polarization modes then results in
\begin{equation}
v_\bot =1 -c^{\text{sc}}_\bot B^2 \sin^2 \theta_B,\quad
v_\| =1-c^{\text{sc}}_\| B^2 \sin^2 \theta_B.\label{eqB4}
\end{equation}
This time, the $\bot$ mode is significantly slower than the $\|$ mode,
since the order of the coefficients is now reversed $c^{\text{sc}}_\bot
>c^{\text{sc}}_\|$.

In a birefringence experiment, the induced ellipticity in the two
cases is different in magnitude as well as in sign. Already at this
stage, we can expect that the same difference will also be visible in
the dichroism. At higher frequencies, the slower mode necessarily has
to exhibit a stronger anomalous dispersion. By virtue of dispersion
relations, we can expect that this goes along with a larger
attenuation coefficient. As a result, the direction of the induced
rotation will be opposite for the two cases, as is confirmed by the
explicit result in Sect.~\ref{sec:MCB}.

\section{Polarization tensors}\label{appB}

The polarization tensor in an external constant magnetic field can be
decomposed into
\begin{equation}
\Pi^{\mu\nu}(k|B) = \Pi_0\, P_{0}^{\mu\nu}+\Pi_\|\, P_\|^{\mu\nu} +
\Pi_\bot\, P_\bot^{\mu\nu},
\label{eqP1}
\end{equation}
where the $P_i$ denote orthogonal projectors, and only the
$\|,\bot$ components are relevant for the dichroism and birefringence
experiments; the corresponding projectors $P_{\|,\bot}$ refer
  to the polarization eigenmodes discussed in the main text
  \cite{Dittrich:2000zu,Gies:1999vb}.  {Dropping terms of higher
  order in the light cone deformation $k^2\simeq 0$ as a
  self-consistent approximation,} the coefficient functions can be
written as
\begin{equation}
\Pi_{\|,\bot}=-\omega^2\sin^2 \theta_B \frac{\alpha}{4\pi}
\left(
  \begin{array}{c}
    -2 \\ 1
  \end{array}
\right)
\int\limits_0^\infty \frac{d s}{s} \int\limits_{-1}^1 \frac{d\nu}{2}
\E^{-\I s \phi_0} \, N_{\|,\bot},\label{111236}
\end{equation}
where the upper component holds for the spinor case and the lower for
the scalar case. The phase reads in both cases
%
\begin{align}\nonumber
\phi_0&=\tm^2 -\omega^2 \sin^2 \theta_B \left( \frac{1-\nu^2}{4}
  -\frac{1}{2} \frac{\cos \nu \te Bs -\cos \te Bs}{\te Bs \sin \te Bs}
  \right)\\
&\simeq \tm^2 +\omega\sin^2\theta_B \frac{(1-\nu^2)^2}{48} \, (\te
  Bs )^2. \label{111235}
\end{align}
%
For completeness, let us list the integrand functions of the spinor
case first,
%
\begin{align}
N_\|^{\text{Dsp}}&
=\frac{\te Bs \cos \nu \te Bs}{\sin \te Bs}\nonumber \\
 &-\te Bs \cot \te Bs \left( 1-
 \nu^2 +\nu \frac{\sin \nu \te Bs}{\sin \te Bs} \right), \nonumber \\
N_\bot^{\text{Dsp}}&
=-\frac{\te Bs \cos \nu \te Bs}{\sin \te Bs} +\frac{\nu \te Bs \,\sin\nu
 \te Bs  \,\cot \te Bs}{\sin \te Bs}  \nonumber \\
&+ \frac{2\te Bs (\cos\nu \te Bs -\cos \te Bs)}{\sin^3 \te Bs}.
 \label{111237}
\end{align}
%
The corresponding lowest-order expansions  in $\tilde{e}Bs$ which are relevant for the
desired approximation are
\begin{align}
N_\|^{\text{Dsp}}&=\frac{1}{2}(1-\nu^2)\left(
1-\frac{1}{3}\nu^2\right)\, (\te Bs)^2,
\nonumber\\
N_\bot^{\text{Dsp}}&= \frac{1}{2}(1-\nu^2)
\left( \frac{1}{2} +\frac{1}{6} \nu^2
\right)\, (\te Bs)^2. \label{InsAbs8}
\end{align}
{Inserting these expansions into \Eqref{111236}, the parameter
  integrations can be performed, resulting in the expressions listed
  in Sect.~\ref{sec:MCF}.} Note that the expansion coefficients in
\Eqref{InsAbs8} also pop up in the final result for the absorption
coefficients and the refractive indices, see below.

The corresponding integrand functions for the scalar case
read\footnote{Compared to \cite{Schubert:2000yt}, we have accounted
  for a global minus sign arising from different global conventions
  for the polarization tensor and the effective action.
}
\cite{Schubert:2000yt}
%
\begin{align}
N_\|^{\text{sc}}=&  -\frac{\te Bs}{\sin \te Bs} \left( -
  \nu^2 +\nu \frac{\sin \nu \te Bs}{\sin \te Bs} \right), \label{eqP2}\\
N_\bot^{\text{sc}}=& +\frac{\nu \te Bs \,\sin\nu
  \te Bs }{\sin^2 \te Bs} \nonumber\\&- \frac{\te Bs}{\sin^3 \te Bs}
 \left( 1+ \cos^2 eBs -2
  \cos \te Bs \cos \nu \te Bs \right). \nonumber
\end{align}
%
The corresponding expansions are
\begin{eqnarray}
N_\|^{\text{sc}}&=&- \frac{1}{2} (1-\nu^2)\, \left(\frac{1}{3} \nu^2
\right) \,(\te Bs)^2,
\label{eqP3}\\
N_\bot^{\text{sc}}&=& -\frac{1}{2} (1-\nu^2) \left(\frac{1}{2} -
  \frac{1}{6} \nu^2\right)\, (\te Bs)^2. \nonumber
\end{eqnarray}
The overall minus sign difference between Eqs.~\eqref{InsAbs8} and
\eqref{eqP3} will be used to cancel the minus sign difference between
the scalar and the spinor case in \Eqref{111236}. Apart from the
overall factor of 2, the desired formulas for the scalar case can be
directly constructed from the spinor case by simple replacements as
suggested by a comparison between Eqs.~\eqref{InsAbs8} and
\eqref{eqP3}.

With the findings of this section, we can directly obtain the
results for the photon absorption coefficients and refractive
indices as given in the main text.

\onecolumngrid
\section{Rotation and Ellipticity at BFRT}\label{appC}

  The BFRT experiment uses a magnetic field with time-varying
  amplitude $B=B_0+\Delta B\cos(\omega_m t+\phi_m)$. The measured
  rotation and ellipticity correspond to the {Fourier
    coefficient of the} light intensity at frequency $\omega_m$.
  {To a good accuracy, the Fourier coefficient can be read off
    from the first-order Taylor expansion of the optical functions
    with respect to $\Delta B$.} The rotation effect for fermionic
  MCPs linear to $\cos(\omega_mt+\phi_m)$ is given by
  Eqs.~(\ref{delthet}) and (\ref{absorption}) for $B=B_0$ and $\chi_0
  = \chi(B_0)$ with
\begin{equation}\label{absorb_linear}
T_{\parallel,\perp}^\text{Dsp} =
\frac{4\sqrt{3}}{\pi\chi_0}
\int\limits_0^1 {\rm d}v\
\frac{\Delta B}{B_0}\left[\left(\frac{4}{\chi_0}\frac{1}{1-v^2}\right)K_{5/3}\left( \frac{4}{\chi_0}\frac{1}{1-v^2}\right)-\frac{2}{3}K_{2/3}\left( \frac{4}{\chi_0}\frac{1}{1-v^2}\right)\right]\times
\frac{\left[ \left( 1-\frac{1}{3}v^2\right)_\parallel,
\left(\frac{1}{2} +\frac{1}{6}v^2\right)_\perp
\right]}{(1-v^2)}.
\end{equation}
The linear term for the ellipticity is given by Eq.~(\ref{psi}) and
(\ref{refraction}) for $B=B_0$ with
\begin{equation}\label{refrac_linear}
I_{\parallel,\perp}^\text{Dsp}\!\!=\!\!2^{\frac{1}{3}}\left(\frac{3}{\chi_0}\right)^{\frac{4}{3}}
\int^{1}_{0} {\rm d}v\,
\frac{2}{3}\frac{\Delta B}{B_0}\left[\tilde{e}^{\prime}_{0}\left[\begin{scriptstyle}-
\left(\frac{6}{\chi_0}\frac{1}{1-v^2}\right)^{\frac{2}{3}}\end{scriptstyle}\right]+\left(\frac{6}{\chi_0}\frac{1}{1-v^2}\right)^{\frac{2}{3}}\tilde{e}^{\prime\prime}_{0}\left[\begin{scriptstyle}-
\left(\frac{6}{\chi_0}\frac{1}{1-v^2}\right)^{\frac{2}{3}}\end{scriptstyle}\right]\right]\times\frac{\left[\left(1-\frac{v^2}{3}\right)_{\parallel},
\left(\frac{1}{2}+\frac{v^2}{6}\right)_{\perp}\right]}{(1-v^{2})^{\frac{1}{3}}}.
\end{equation}
The corresponding equations in the case of scalar MCPs are analogous.
\twocolumngrid

\end{document}